\documentclass{elsart}

\usepackage{natbib}
\usepackage{graphicx}
\usepackage{amssymb}

\journal{New Astronomy}

\begin{document}

\def\lsim{\lower.5ex\hbox{$\; \buildrel < \over \sim \;$}}
\def\gsim{\lower.5ex\hbox{$\; \buildrel > \over \sim \;$}}
\def \simeq{\lower.3ex\hbox{$\; \buildrel \sim \over - \;$}}
\def\eg{{\it e.g.,} }
\def\etal{{\em et al.} }
\def\ie{{\em i.e.,} }

\begin{frontmatter}

\title{Mass loss from viscous advective disc}

\author{Indranil Chattopadhyay \corauthref{cor}}
\address{Department of Astronomy and Space Science, Chungnam National Univ,
Daejeon, South Korea}
\ead{indra@canopus.cnu.ac.kr}
\corauth[cor]{Corresponding author.}

\author{Santabrata Das}
\address{ARCSEC, Sejong University, Seoul, South Korea}
\ead{sbdas@canopus.cnu.ac.kr}

\begin{abstract}
Rotating transonic flows are long known to admit standing or oscillating
shocks. The excess thermal energy in the post shock flow drives
a part of the in falling matter as bipolar outflows.
We compute mass loss from a viscous advective disc.
We show that the mass outflow
rate decreases with increasing viscosity of the accretion disc, since
viscosity weakens the centrifugal barrier that generates the shock.
We also show that the optical depth of the post-shock matter decreases
due to mass loss which may soften the spectrum from such a mass losing disc.
\end{abstract}


\end{frontmatter}

\section{Introduction}

Accretion onto black holes has been intensely studied for the last
three decades, in order to explain observed
luminosities of galactic black hole
candidates and AGNs, their spectral states, and 
the formation of jets or outflows around black hole candidates. 
Since black holes have neither hard surface, nor
intrinsic atmosphere from which
outflows may originate, jets/outflows has to originate from the
accreting matter. 
Recently, \citet{gal03} showed that jets originate 
from accretion discs which are in hard
spectral states. 
This fact suggests that the origin of jet is strongly correlated with
the spectral state of the accretion disc, and therefore with the accretion
process itself.

The unique inner boundary condition for accretion
onto a black hole, necessarily demands that the black hole accretion
is globally transonic. 
\citet{c89} showed that transonic matter
suffers centrifugal pressure mediated shock, and
the shock is located further out
as the specific energy and the specific angular momentum
of the flow is increased.
Such a shocked accretion disc model was used to compute the spectral states
of the black hole candidates \citep{ct95,cm06},
where the post-shock tori --- CENBOL (CENtrifugal Pressure
supported BOundary Layer), produces the hard power-law tail by
inverse-Comptonizing the softer photons from the outer disc.
Moreover, it has also been shown, both numerically \citep{mlc94,mrc96}
and analytically \citep{c99,dc99,dcnc01},
that the extra thermal gradient force in the CENBOL drives a significant portion
of accreting matter as bipolar outflows.
In particular, \cite{dcnc01} showed that outflows from CENBOL
are generated in hard spectral states of accretion discs.
These bipolar outflows are
believed to be the precursor of astrophysical jets.

So far, there is no theoretical attempt to
compute mass outflow
rates from a viscous advective disc.
As viscous flow approaches the black hole, the local angular momentum
of the flow decreases
while the local energy increases.
Thus, in a viscous flow there are two competing processes to
determine the shock formation.
Should the shock move inwards as we increase the
viscosity for flows with identical outer boundary conditions
or vice-versa?
How would the position of shock affect the mass outflow rates?
We are going to address the above mentioned issues in the present paper.

In the next section, we present the model assumptions and the
equations of motion of the disc-jet system. In \S 3, we present
the solution procedure. In \S 4, we present the results,
and in the last section we draw concluding remarks.

\section{Model Assumptions and Equations of motion}
We begin with a steady, viscous, axisymmetric accretion 
flow on to a Schwarzschild black hole. The space-time 
property is approximated
by the Paczy{\'n}ski-Wiita potential 
\citep{pw80}.
In addition, jets are considered to be inviscid.
Outflows are supposed to be lighter
than the accretion disc, and definitely of lower angular momentum
than at least the outer edge of the disc. Consequently, the differential
rotation is expected to be much less than that in the accretion disc.
Therefore, viscosity in jets can be ignored.
In Fig. 1, a schematic diagram of an advective accretion disc-jet
system 
is presented. In this figure the pre-shock flow,
the shock location ($x_s$), the post-shock flow (abbreviated as CENBOL)
and the jet between the funnel wall (FW) and centrifugal barrier (CB)
have been marked. We define FW and CB in \S 2.2.
Furthermore, we use a system of units
such that $2G=M_{\rm BH}=c=1$, where
$G$, $M_{\rm BH}$, and $c$ are the gravitational constant, the
mass of the black hole, and the velocity of light, respectively.

\begin{figure}

\begin{center}
\includegraphics[scale=0.55]{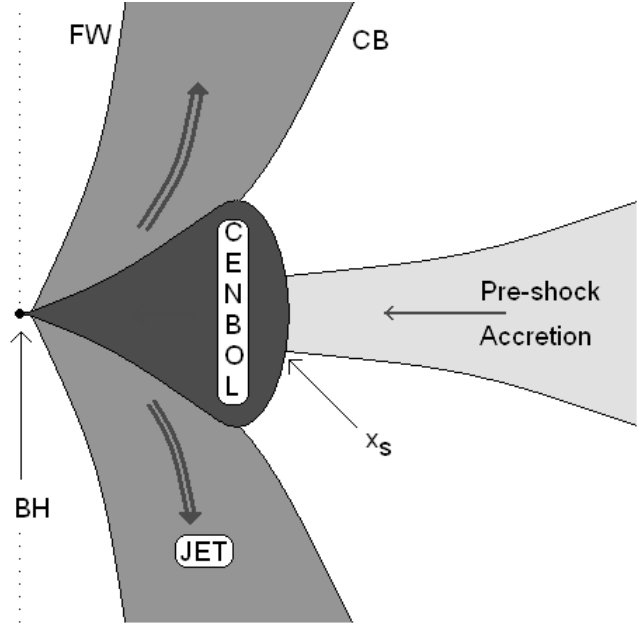}
\end{center}

\caption{ Schematic diagram of disc-jet system (with up-down symmetry)
where the position 
of black hole (BH), shock location ($x_s$),
CENBOL, centrifugal barrier (CB), funnel wall (FW) etc are shown.
The dotted line is the axis of symmetry. 
}
\label{fig1}
\end{figure}
\subsection{Equations for accretion}
In the steady state, the dimensionless hydrodynamic equations
of motion for accretion are \citep{gut},

\noindent the radial momentum equation :
$$
u \frac {du}{dx}+\frac {1}{\rho}\frac {dP}{dx}
-\frac {\lambda^2(x)}{x^3}+\frac {1}{2(x-1)^2}=0,
\eqno(1a)
$$
\noindent the baryon number conservation equation :
$$
\dot M = 2 \pi \Sigma u x,
\eqno(1b)
$$
\noindent the angular momentum conservation equation :
$$
u \frac {d\lambda(x)}{dx}+\frac{1}{\Sigma x}
\frac {d}{dx}\left( x^2 W_{x\phi}\right)=0,
\eqno(1c)
$$
and finally, the entropy generation equation :
$$
u T \frac {ds}{dx}= -\frac{{\alpha}_{\Pi}}{\gamma}
x(ga^2+\gamma u^2)\frac{d\Omega}{dx},
\eqno(1d)
$$
where, variables $u$, $a$, $\rho$, $P$ and $\Omega(x)$, and $\lambda(x)$ in the 
above equations are the radial velocity, sound speed, density, isotropic
pressure, angular velocity and specific angular momentum of the flow
respectively. Here $\Sigma$ is vertically integrated density,
$W_{x\phi}$($=- \alpha_{\Pi}\Pi$, $\Pi$
is the vertically integrated total pressure, $\alpha_{\Pi}$
is the viscosity parameter) denote the 
viscous 
stress, $s$ is the specific entropy of the flow, and $T$ is the
local temperature.
The disk 
is assumed to be in hydrostatic equilibrium in the 
vertical direction.
Furthermore, $g=I_{n+1}/I_n$, $n=1/(\gamma -1)$ and
$I_n=(2^nn!)^2/(2n+1)!$ \citep{matetal84},
and the adiabatic index $\gamma=4/3$ is used through out the paper.

\subsection{Equations of motion for outflows}
\begin{figure}

\begin{center}
\includegraphics[scale=0.4]{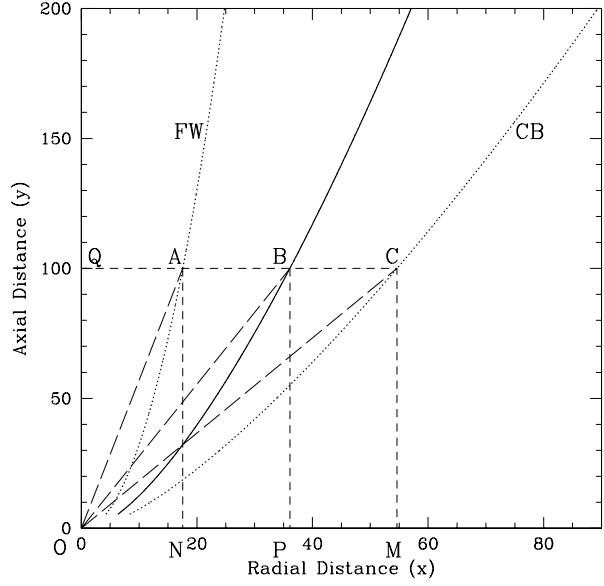}
\end{center}

\caption{ Jet geometry for $\lambda =1.75$. In figure $OA=r_{FW}$,
$OC=r_{CB}$, $QC=x_{CB}$, $CM=y_{CB}$; $CM=BP(y_j)=AN(y_{FW})$;
$QA=x_{FW}$; $QB=x_j$;
And $OB=r_j$. FW and CB are marked in the figure.
}
\label{fig2}
\end{figure}
Numerical simulations by \citet{mrc96} suggests that
the outflowing matter tends to emerge out between
two surfaces namely, the funnel wall (FW)
and the centrifugal barrier (CB).
In Fig. 2, the schematic diagram of the jet geometry is shown.
Geometric surfaces CB and FW are marked in the figure.
The centrifugal barrier (CB) surface is defined as the pressure maxima surface
and is expressed as
$$
x_{CB}=\left[ 2 \lambda^2 r_{CB} (r_{CB}-1)\right]^{1/4},
\eqno(2a)
$$
where, $r_{CB}=\sqrt{x^2_{CB}+y^2_{CB}}{\equiv}$ spherical radius of
CB. Here $x_{CB}$, $y_{CB}$ are the cylindrical radius and axial
coordinate (\ie height at $r_{CB}$) of CB.
We compute the jet geometry with respect to $y_{CB}$ \ie
$y_{FW}=y_j=y_{CB}$,
where $y_{FW}$ and $y_j$ are the height of FW and the jet
at $r_{CB}$, respectively.
The FW is obtained by considering null effective potential
and is given by
$$
x^2_{FW}=\lambda^2 \frac{(\lambda^2-2)+
\sqrt{(\lambda^2-2)^2-4(1-y^2_{CB})}}{2},
\eqno(2b)
$$
where, $x_{FW}$
is the
cylindrical radius of FW.
Using Eqs. (2a) and (2b), we estimate the cylindrical
radius of the outflow
$$
x_j=\frac{x_{FW}+x_{CB}}{2}.
\eqno(2c)
$$
The spherical radius of the jet is given by
$r_j=\sqrt{x^2_j+y^2_j}$.
In Fig. 2, $OB$ ($=r_j$) defines the streamline (solid) of the outflow.
The total area function of the jet is obtained as,
$$
{\mathcal A=2\pi (x^2_{CB}-x^2_{FW})}
\eqno(2d)
$$
The integrated momentum balance equation for jet is given by,
$$
{\mathcal E}_j= \frac{1}{2}v^2_j+na^2_j +\frac {\lambda^2_j}{2x^2_j}
-\frac{1}{2(r_j-1)},
\eqno{(3a)}
$$
where ${\mathcal E}_j$ and $\lambda_j$
are the specific energy and the angular momentum of the jet, respectively.
The integrated continuity equation is,
$$
{\dot M}_{\rm out}=\rho_j v_j {\mathcal A},
\eqno{(3b)}
$$
and instead of the entropy generation equation we have the polytropic equation
($p_j=K_j\rho^{\gamma}_j$) of state for the jet.
In Eqs. (3a-3b), the suffix `$j$' indicates jet variables,
where $v_j$, $a_j$, and $\rho_j$ are the velocity, the sound speed and the
density of the jet.

\section{Method}

Since post-shock disc is the source of outflows, we are interested in
only those accretion solutions which includes stationary shocks.
A shocked accretion solution has two X-type critical points ---
inner ($x_{ci}$) and outer ($x_{co}$) critical points. Critical points
are radial distances at which the gradient of velocity \ie
$du/dx{\rightarrow}0/0$. The accretion solution is solved by
supplying $x_{ci}$ and angular momentum $\lambda_i$ at $x_{ci}$ and integrating
the equations once from $x_{ci}$ inwards and then outwards [as in
\citep{cd04}].
The conditions for steady shock require conservation of energy flux,
mass flux,
and momentum flux across the shock front [\eg \citep{ll59}] and are
given by,
$$
\left[{\mathcal E} \right]=0; ~~~~
\left[\dot{M} \right]=0; ~~~~
\left[{\Pi}\right]=0.
\eqno{(4a)}
$$
In presence of mass loss, the condition for conservation of mass
flux takes the following form,
$$
{\dot{M}}_{+} ={\dot {M}}_{-}-{\dot {M}}_{\rm out}
={\dot {M}}_{-}(1-R_{\dot m}),
\eqno{(4b)}
$$
where, subscripts ``$-$'' and ``$+$'' refer
to quantities before and after the shock, respectively.
The mass outflow rate $R_{\dot m}$ 
is the ratio between mass flux of the outflow (${\dot {M}}_{\rm out}$)
and the pre-shock
accretion rate (${\dot {M}}_{-}$).
Using the above shock conditions,
the pre-shock sound speed and bulk velocity can be expressed as,
$$
a^2_-=\frac{C_1u_-}{g}-\frac{\gamma u^2_-}{g},
\eqno{(5a)}
$$
and,
$$
\left[1-\frac{2\gamma}{g(\gamma -1)}\right]u^2_-+\frac{2C_1}{g(\gamma -1)}u_-
+\left[\frac{\lambda^2_s}{x^2_s}-\frac{1}{x_s-1}-2{\mathcal E}_s \right]=0
\eqno{(5b)}
$$
where $C_1=(1-R_{\dot m})(ga^2_++\gamma v^2_+)/u_+$,
and $({\mathcal E}_s, \lambda_s)$
are the energy and the angular momentum at the shock.
As we integrate from $x_{ci}$ outwards, Eqs. (5a-b) are used to
compute pre-shock flow variables. These pre-shock variables
(\ie $u_-$, $a_-$) are then used to find $x_{co}$.
However,
Eqs. (5a-b) show that the information of mass loss is in the
shock condition itself. This implies that, in order to obtain
shock the accretion-jet equations
have to be solved simultaneously.

Jet equations are solved by employing the so-called `critical
point analysis' \citep{ttaf}.
We differentiate Eqs. (3a-3b) with respect to $r(=r_{CB})$,
and obtain the expression for $dv_j/dr$.
The critical point ($r_{jc}$) condition for jet is,
$$
v^2_{jc}=a^2_{jc}=\left[\frac{1}{2(r_{jc}-1)^2}\left(\frac{dr_j}{dr}\right)_{r_c}
-\frac{\lambda^2}{x^3_{jc}}\left(\frac{dx_j}{dr} \right)_{r_c}\right]
\left[\frac{1}{{\mathcal A}_c}\left(\frac{d{\mathcal A}}{dr}\right)_{r_c}
\right]^{-1},
\eqno{(3c)}
$$
Once the flow variables at the jet critical points are known, we
integrate the jet equations from $r_{c}$ to the jet base ($x_s$).

The expression for $R_{\dot m}$ is calculated by assuming
that jets are launched with the same density as the post-shock flow and is
given by,
$$
R_{\dot m} =\frac{{\dot M}_{\rm out}}{{\dot M}_-}
=\frac{Rv_{j}(x_s) {\mathcal A}(x_s)}
{4 \pi \sqrt{\frac{2}{\gamma}}x^{3/2}_s (x_s-1)a_+u_{-}},
\eqno{(6)}
$$
where the compression ratio is expressed as
$R={\Sigma_+}/{\Sigma_-}$.
The self-consistent method to obtain shocks in presence of
mass loss is the following : 
initially the mass loss is not considered, and the shock is found out for
accretion flows. Using
$({\mathcal E}_j,\lambda_j)=({\mathcal E}_s,\lambda_s)$
as inputs, we solve the jet equations
and calculate the values of jet variables
at $r=x_s$. Once we obtain the value of $v_j$, ${\mathcal A}$
at $r=x_s$, we use Eq. (6) to calculate $R_{\dot m}$. The
value of $R_{\dot m}$ is used
in Eqs. (5a-5b), and the new shock
location is obtained. We iterate this process till the solutions converge.

\begin{figure}
\begin{center}
\includegraphics[scale=0.5]{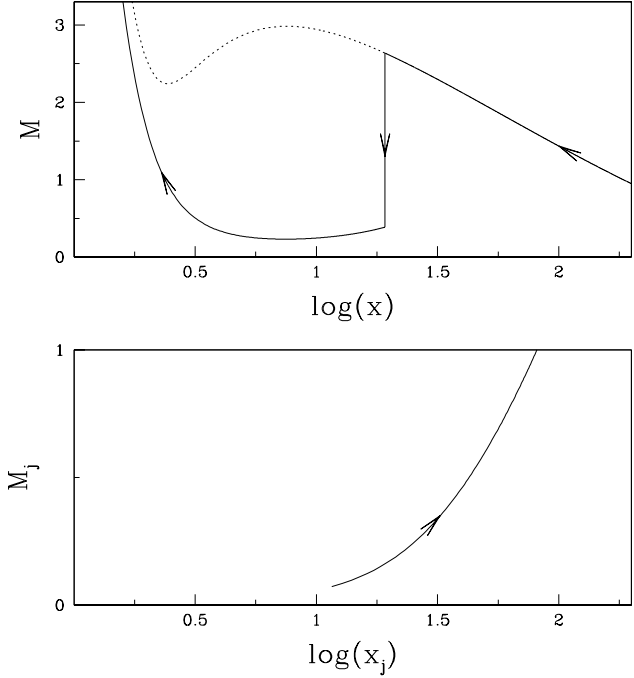}
\end{center}
\caption{Upper panel: Inflow Mach number ($M=u/a$) with $log(x)$.
The inflow parameters are
$x_{ci}=2.445$, $\lambda_i=1.75$, and
$\alpha_{\Pi}=0.002$, where, $x_s=19.21$, ${\mathcal E}_s=0.0014$,
$\lambda_s=1.756$ $x_{co}=206.49$, $\lambda_o=1.772$. The dotted curve
is the shock free solution.
Lower panel: Outflow Mach number
($M_j=v_j/a_j$) with $log(x_j)$,
the outflow critical point $x_{jc}=81.26$ ($r_{jc}=
346.18$), and the jet coordinates at the base
is given by $x_{jb}=10.27$ ($r_{jb}=16.62$).
The mass loss rate is $R_{\dot m}=0.055$.
}
\label{fig3}
\end{figure}

\section{Results}

Supersonic matter, which was subsonic at the outer edge of the disc,
suffers a discontinuous transition due to centrifugal barrier at shock ($x_s$)
and becomes sub-sonic.
A significant part of the  post-shock matter is ejected
as bipolar outflows due to excess thermal pressure
at the shock and the rest enters into the black hole through
$x_{ci}$.
In Fig. 3, we present such a global inflow-outflow
solution. In the top panel
the Mach number of the accretion flow is plotted
with $log(x)$. The solid curve represents shock induced accretion
solution.
The inflow parameters are $x_{ci}= 2.445$ , $\lambda_i=1.75$,
and $\alpha_{\Pi}=0.002$. 
In the lower panel, the outflow Mach number
$M_j$ is plotted
with $log(x_j)$. In presence of mass loss ($R_{\dot m}=0.055$)
the shock forms at $x_s=19.21$ (vertical line in the top panel),
and the outflow
is launched with energy and angular momentum at the shock
(${\mathcal E}_s, \lambda_s =0.0014, 1.756$).
The outflow is plotted up to its sonic point ($x_{jc}=81.26$).

\begin{figure}
\begin{center}
\hskip 7.5cm \includegraphics[scale=0.6]{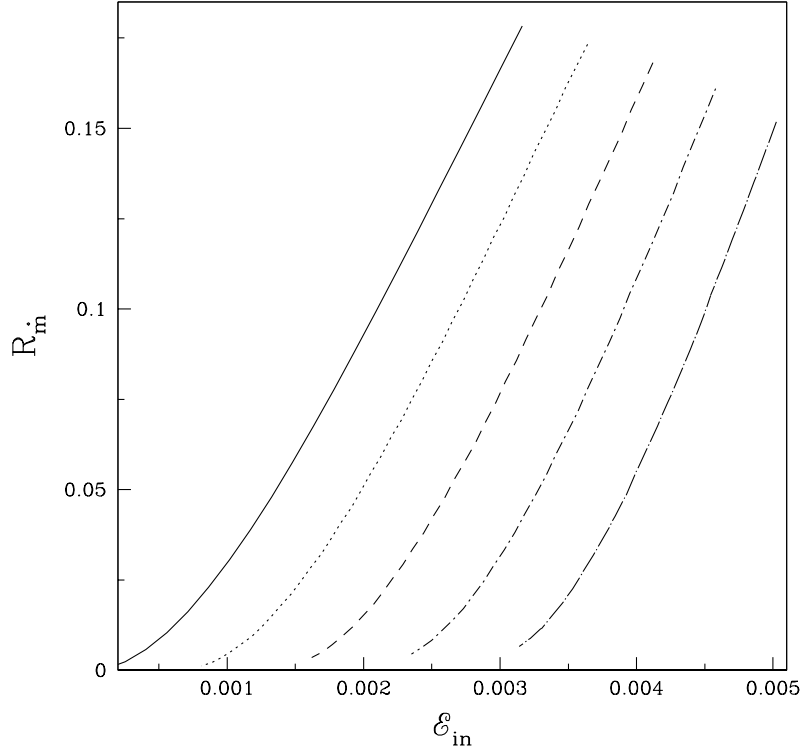}
\end{center}
\vskip 0.5cm
\caption{Variation $R_{\dot m}$ with ${\mathcal E}_{\rm in}$, for
${\alpha}_{\Pi}=0$ (solid), $0.005$ (dotted), $0.01$ (dashed),
and $0.015$ (dashed-dotted), $0.02$ (long-dashed) respectively,
and $\lambda_i=1.75$.
}
\label{fig4}
\end{figure}

We now present how the mass outflow rate is affected by viscosity
in the disc.
In Fig. 4, $R_{\dot m}$ is plotted with ${\mathcal E}_{\rm in}$
(the energy at $x_{ci}$),
for a set of ${\alpha}_{\Pi}$.
For instance,
solid curve (${\alpha}_{\Pi}=0$) represents
mass outflow rates for inviscid accretion.
Other curves are obtained for
the following values of ${\alpha}_{\Pi}$:
$0.005$ (dotted), $0.01$ (dashed),
and $0.015$ (dashed-dotted), and $0.02$ (long dash-dotted) respectively.
All the curves are drawn for $\lambda_i=1.75$.
We observe that higher energetic flow produces higher
$R_{\dot m}$ for a given ${\alpha}_{\Pi}$.
Conversely, for the same value of ${\mathcal E}_{\rm in}$
if $\alpha_{\Pi}$ is increased,
then $R_{\dot m}$ is decreased significantly.
It is to be noted that,
for each ${\alpha}_{\Pi}$ there is a cut-off in $R_{\dot m}$ at
the higher energy
range, since standing shock conditions are not satisfied there.
Non-steady shocks may still form in those regions, but investigation
of such phenomena is beyond the scope of this paper.

\begin{figure}

\vskip -1.0cm
\hskip -1.5cm
\includegraphics[scale=0.31]{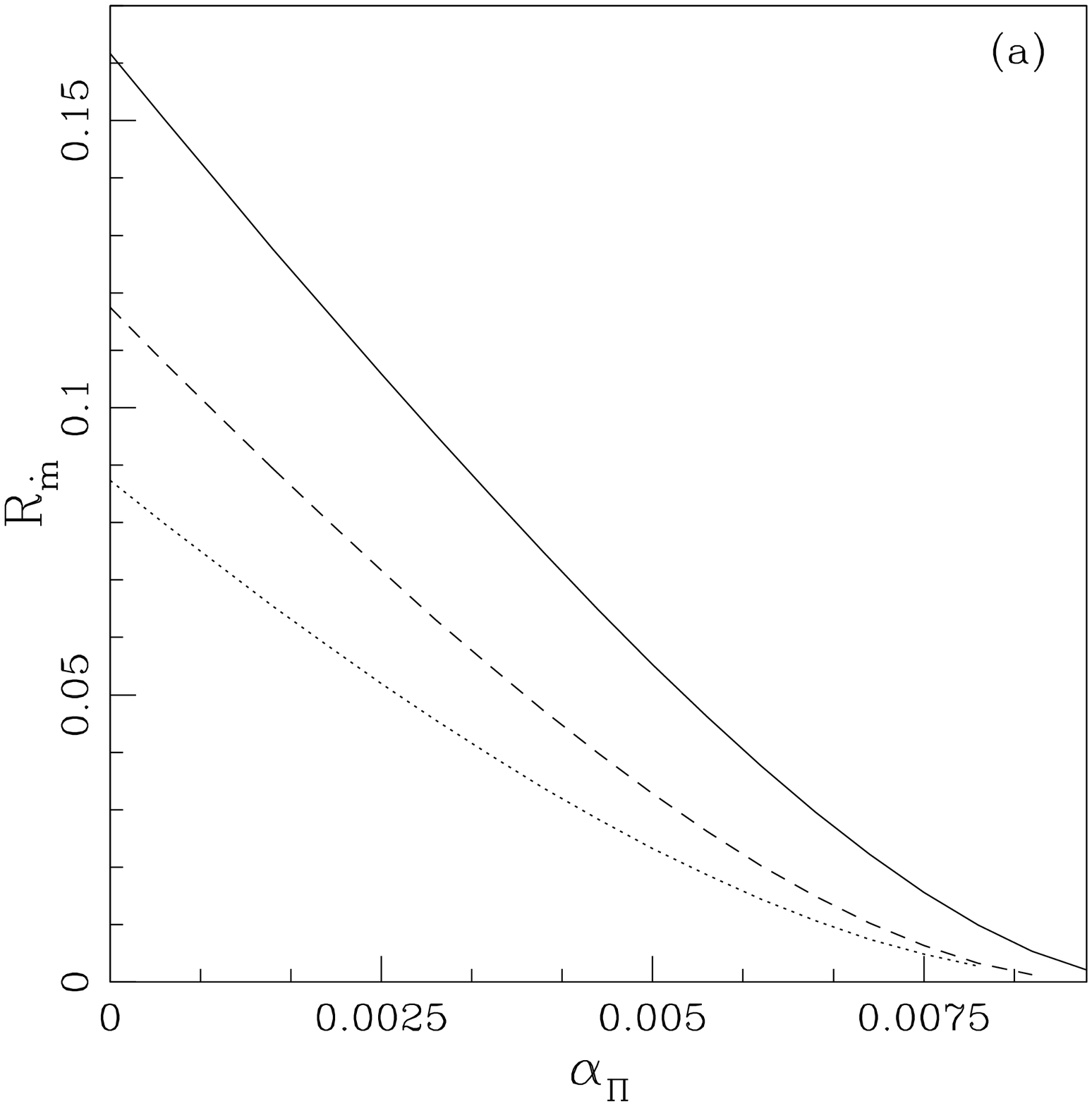}

\hskip -1.5cm
\includegraphics[scale=0.31]{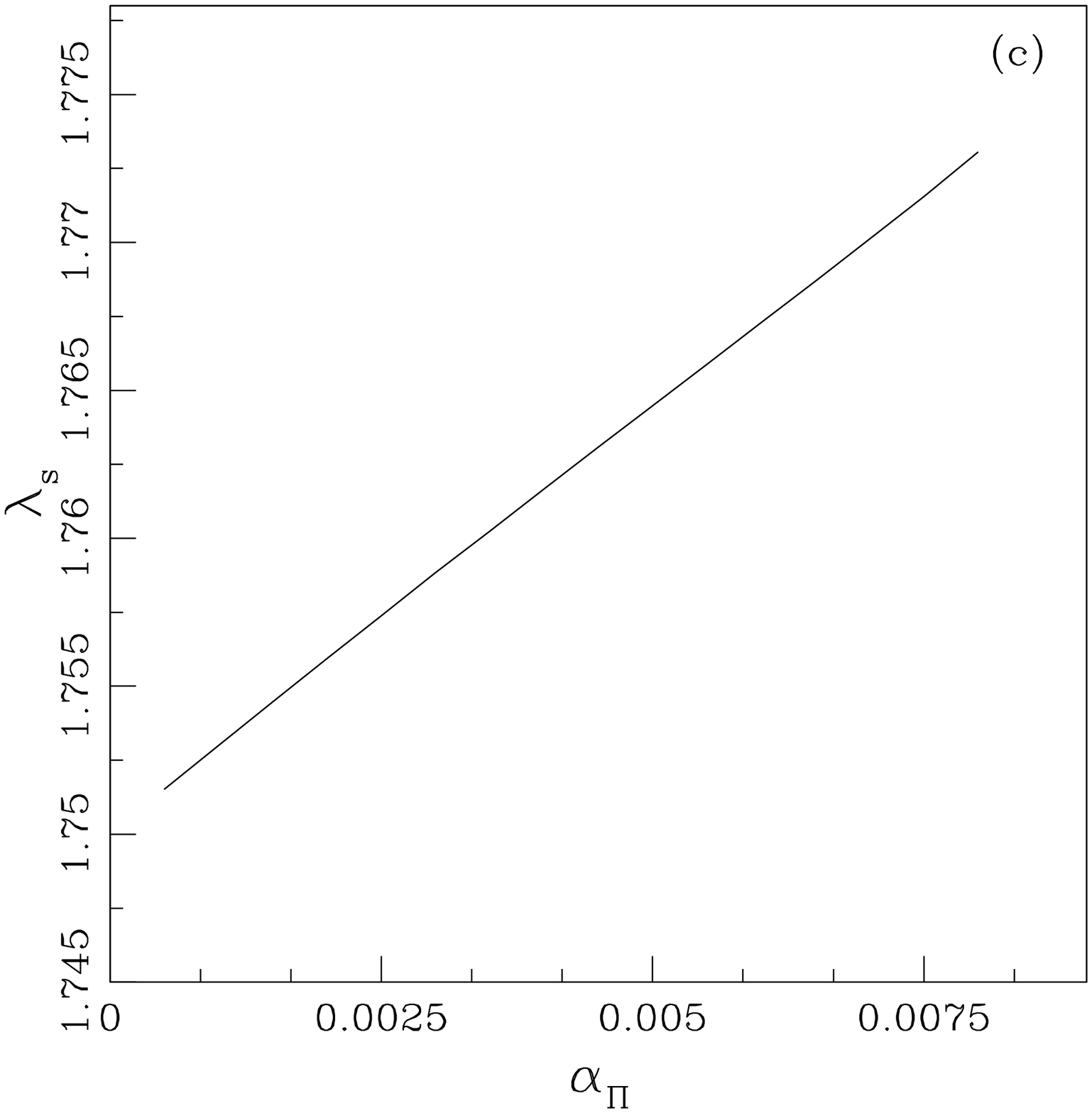}

\hskip -1.5cm
\includegraphics[scale=0.31]{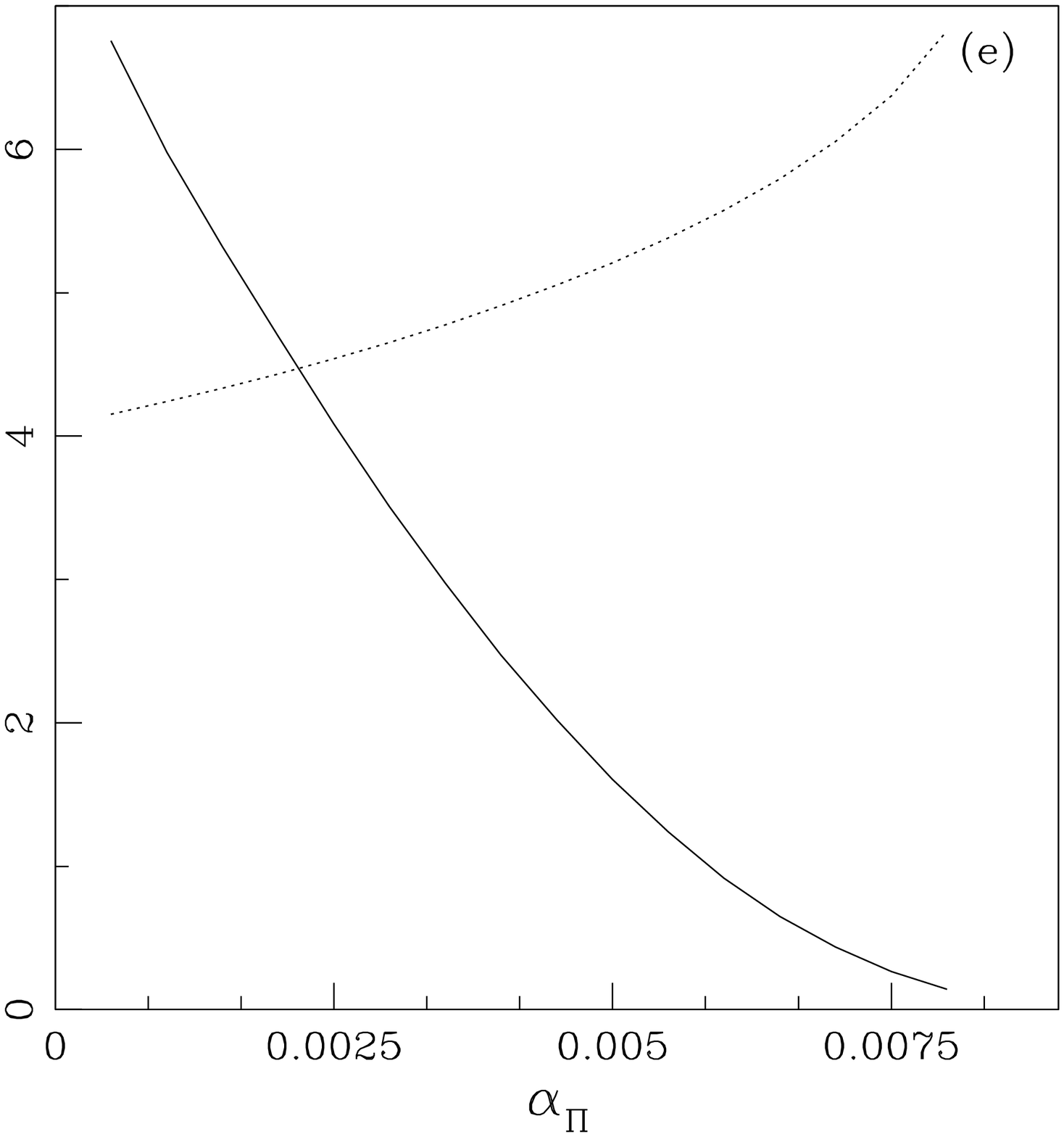}

\vskip -19.0cm
\hskip 7.5cm
\includegraphics[scale=0.31]{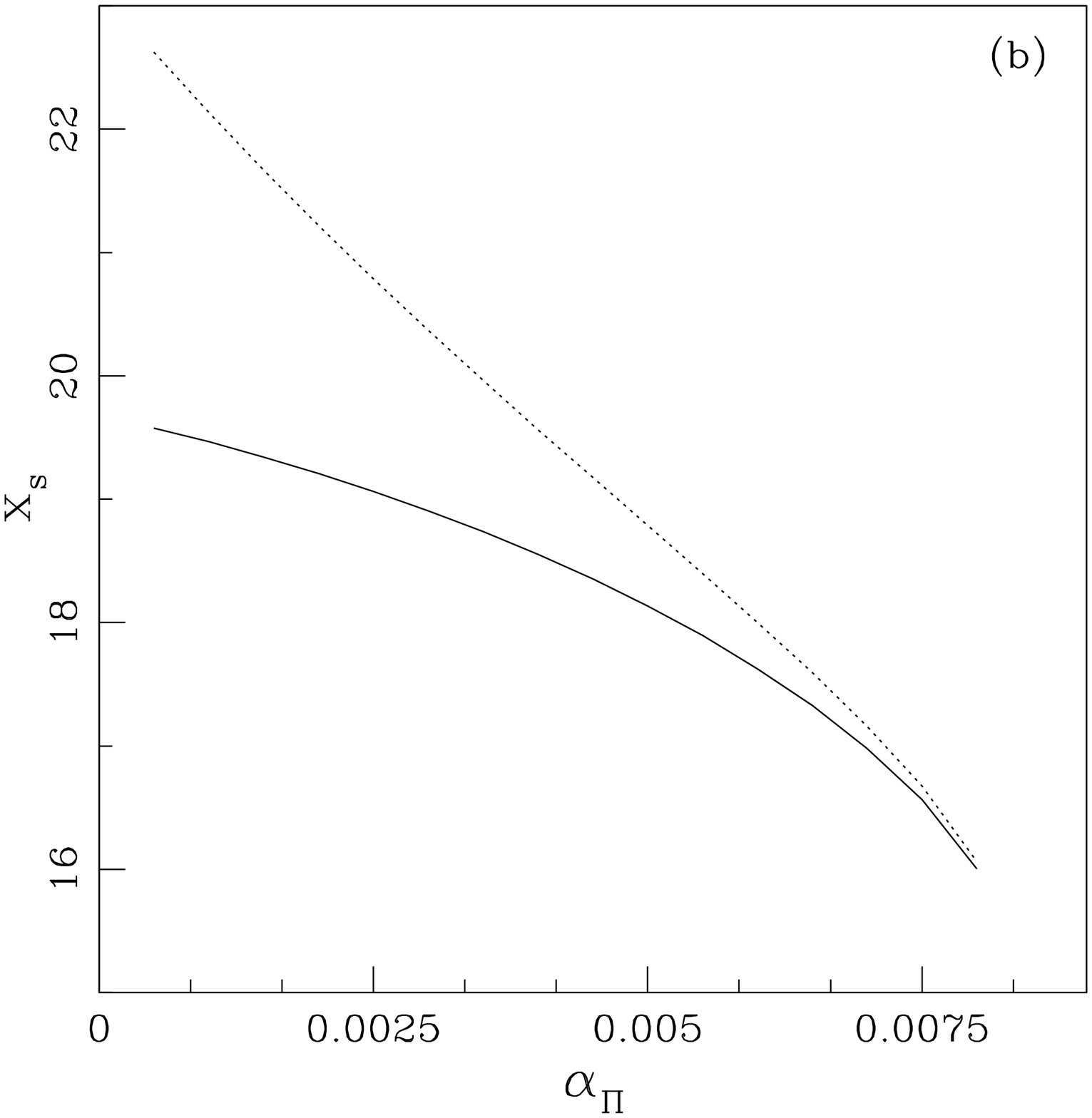}

\vskip 0.0cm
\hskip 7.5cm
\includegraphics[scale=0.31]{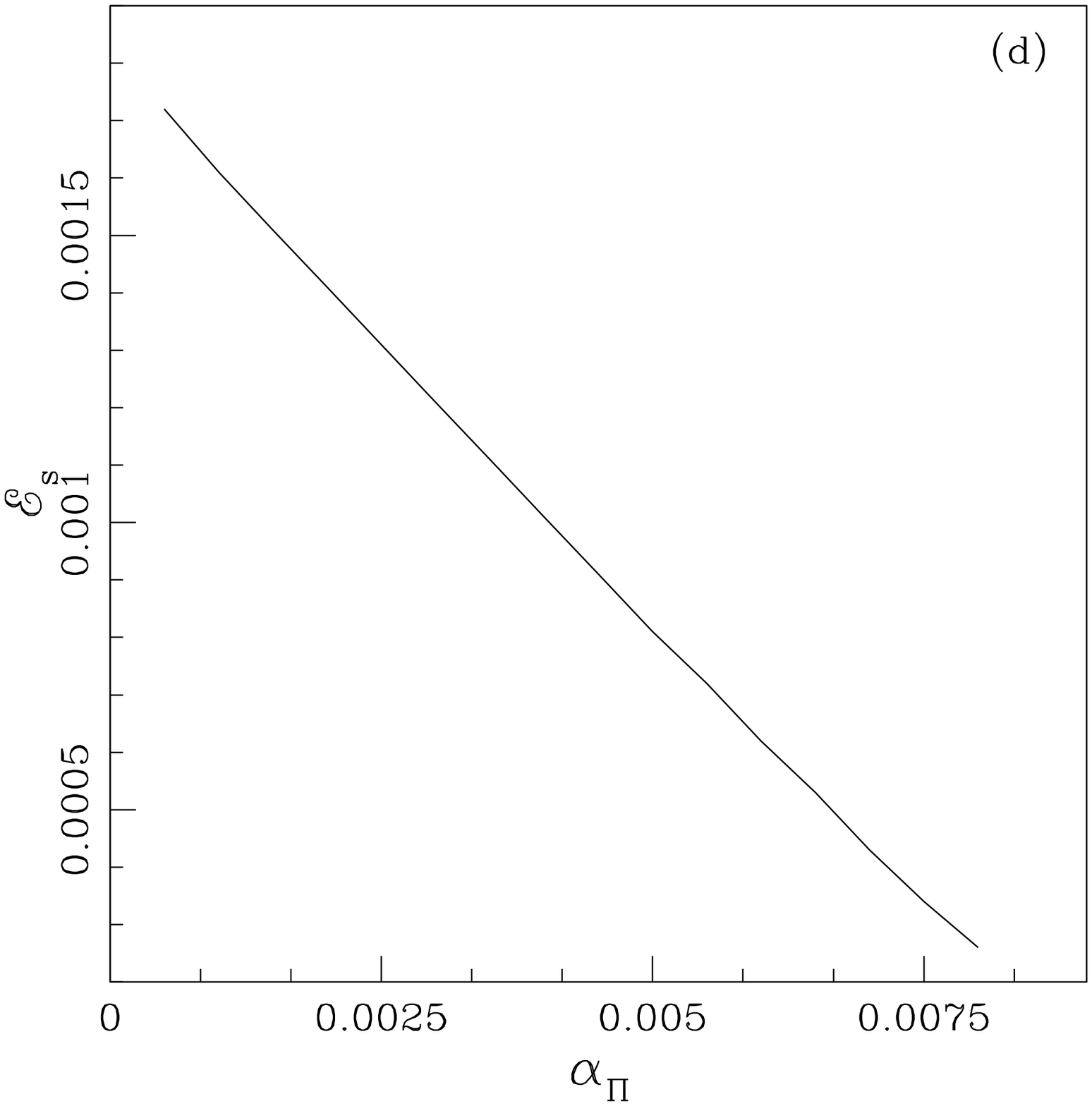}

\hskip 7.5cm
\includegraphics[scale=0.31]{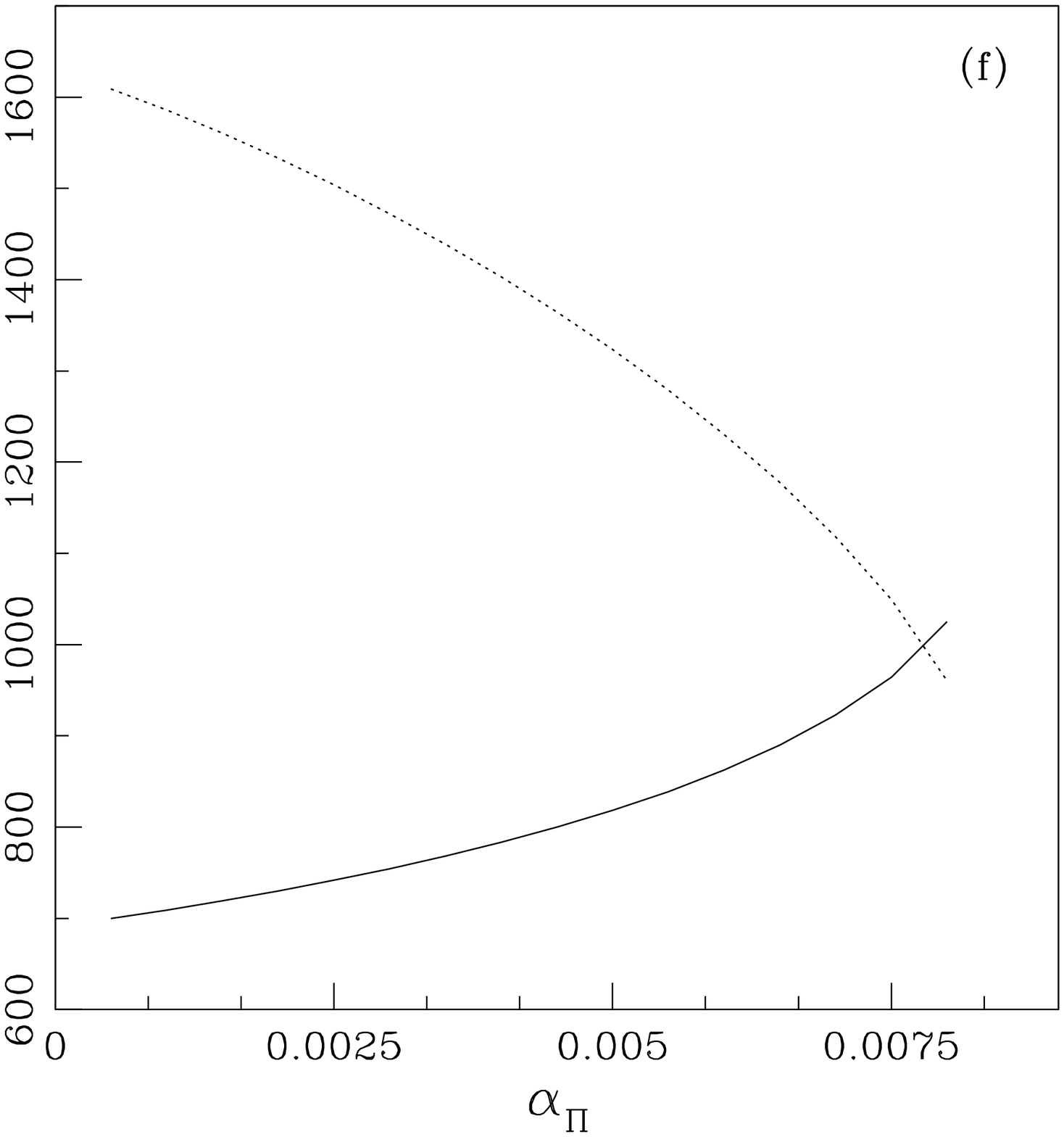}

\caption{(a) $R_{\dot m}$ vs $\alpha_{\Pi}$, for
for $\lambda_i=1.8$ (solid), $1.775$ (dashed) and $1.75$
(dotted), respectively. Inner sonic points are
$x_{ci}=2.313$, $2.375$, and $2.445$, respectively.
(b) Variation of $x_s$ with $\alpha_{\Pi}$, $\lambda_i=1.75$
$x_{ci}=2.445$.
Dotted curve represents solution without mass loss, and solid
represents $x_s$ with mass loss.
(c) Variation $\lambda_s$ with $\alpha_{\Pi}$,$\lambda_i=1.75$
$x_{ci}=2.445$.
(d) Variation ${\mathcal E}_s$ with $\alpha_{\Pi}$,$\lambda_i=1.75$
$x_{ci}=2.445$. (e) Variation of $R$ (dotted) and
$v_j{\mathcal A}$ (solid) with $\alpha_{\Pi}$, $\lambda_i=1.75$,
$x_{ci}=2.445$.
(f) Variation of $x^{3/2}_s(x_s-1)$ (dotted) and $a_{s+}u_-{\times}
5e+4$ (solid)
with $\alpha_{\Pi}$, $\lambda_i=1.75$, $x_{ci}=2.445$.  
}
\label{fig5}
\end{figure}

In Fig. 5a, $R_{\dot m}$ is plotted with $\alpha_{\Pi}$, for
$(x_{ci},\lambda_i)=(2.313,1.8)$
(solid),
$(2.375,1.775)$ (dashed) and $(2.445,1.75)$ (dotted), respectively.
For a given $\alpha_{\Pi}$, $R_{\dot m}$ increases with
higher $\lambda_i$. This suggests that outflows are primarily
centrifugal pressure driven.
However, $R_{\dot m}$ decreases drastically with $\alpha_{\Pi}$
for each of the curves.
The question however is, why $R_{\dot m}$ decreases with $\alpha_{\Pi}$?
We wish to discuss the underlying physics and consider a representative case
[\eg $(x_{ci},\lambda_i)=(2.445,1.75)$]. 
In Fig. 5b, shock location ($x_s$) is plotted with $\alpha_{\Pi}$,
for solutions without mass loss (dotted) and with mass loss (solid).
As viscosity is increased the shock location is decreased.
Generally, in presence of mass loss shock forms closer to the black hole.
Mass loss reduces the post-shock pressure, and therefore the shock
moves inwards to maintain pressure balance across it.
However, the two curves tend to merge at higher $\alpha_{\Pi}$,
which signifies that the mass loss decreases with $\alpha_{\Pi}$.
In Figs. 5(c-d), $\lambda_s$ and ${\mathcal E}_s$ are plotted
with $\alpha_{\Pi}$,
respectively.
In this case, $\lambda_i$ is same, therefore
with increasing $\alpha_{\Pi}$ there is a moderate increase
of $\lambda(x)$ as we integrate outwards,
however, ${\mathcal E}(x)$ decreases sharply, which results in the
lower value of $x_s$.
The fractional increase of $\lambda_s$ is ${\sim}1.1\%$.
The fractional decrease of ${\mathcal E}_s$ is ${\sim}8.2\%$.
As $x_s$ moves inwards, the base area ${\mathcal A}|_{r=x_s}$
of the jet decreases, and consequently the
base velocity ($v_j|_{r=x_s}$)
of the jet also decreases,
while the compression ratio ($R$) increases. 
In Fig. 5e, $v_j{\mathcal A}$ (solid) and $R$ (dotted) is plotted with
$\alpha_{\Pi}$. The quantity $v_j{\mathcal A}$ decreases by $90\%$, while $R$
increases by mere $60\%$.
From Eq. (6), we see that in the denominator of the expression
for $R_{\dot m}$,
there is a term $x^{3/2}_s(x_s-1)
a_+u_-$. In Fig. 5f, both the quantities $x^{3/2}_s(x_s-1)$ (dotted) and
$a_+u_-$ (solid) is plotted with $\alpha_{\Pi}$. The quantity
$x^{3/2}_s(x_s-1)$ decreases by $37\%$ and $a_+u_-$ (multiplied by a factor
$5{\times}10^4$ in the figure) increases by
$42\%$. Therefore, as $x_s$ decreases, the decrease of $v_j{\mathcal A}$
dominates all other quantities in Eq. (6), and hence $R_{\dot m}$ decreases
with the increasing $\alpha_{\Pi}$.
In Figs. 5(b-f), we have fixed
($\lambda_i,x_{ci}$), and increased $\alpha_{\Pi}$ to see
how viscosity affects shock and consequently $R_{\dot m}$.
If one increases $\alpha_{\Pi}$ for fixed ($\lambda_i,x_{ci}$), then
${\mathcal E}_{\rm in}$ will decrease anyway [\eg Fig. (1c)
of Chakrabarti \& Das 2004], hence ${\mathcal E}_s$ will
be lower for higher $\alpha_{\Pi}$. In other words, we have studied
less energetic accretion flows though we have increased the
$\alpha_{\Pi}$ parameter, and saw that for such flows increasing
viscosity parameter decreases the shock location and therefore
the mass outflow rate. The question however is,
what happens to flows starting with same outer boundary conditions,
as viscosity is increased? 

In Fig. 6, the variation of $M$ (upper panel), $\lambda$
(middle panel), and
${\mathcal E}$ (lower panel) with $log(x)$ is presented.
The flow is launched at the outer edge ($x_{\rm inj}=200$)
with energy and angular momentum corresponding to
$({\mathcal E}_{\rm inj},
\lambda_{\rm inj})=(0.002,1.75)$.
For inviscid case, the shock forms at $x_s=20.3$ (dashed) and 
mass outflow rate is calculated to be $R_{\dot m}=0.093$.
As viscosity is
incorporated (${\alpha}_{\Pi}=0.003$; dotted), the shock forms at
$x_s=11.65$ and the outflow rate is reduced to $R_{\dot m}=0.07$.
As viscosity is enhanced,
${\mathcal E}(x)$ is increased and $\lambda(x)$ is decreased
along the flow.
At the shock, the effect of 
energy increment, cannot compensate the effect due
to angular momentum reduction.
This weakens the centrifugal barrier and causes the shock to move inward.
As shock moves inwards the jet-base area
decreases, which reduces the mass outflow rate.
\begin{figure}

\begin{center}
\includegraphics[scale=0.6]{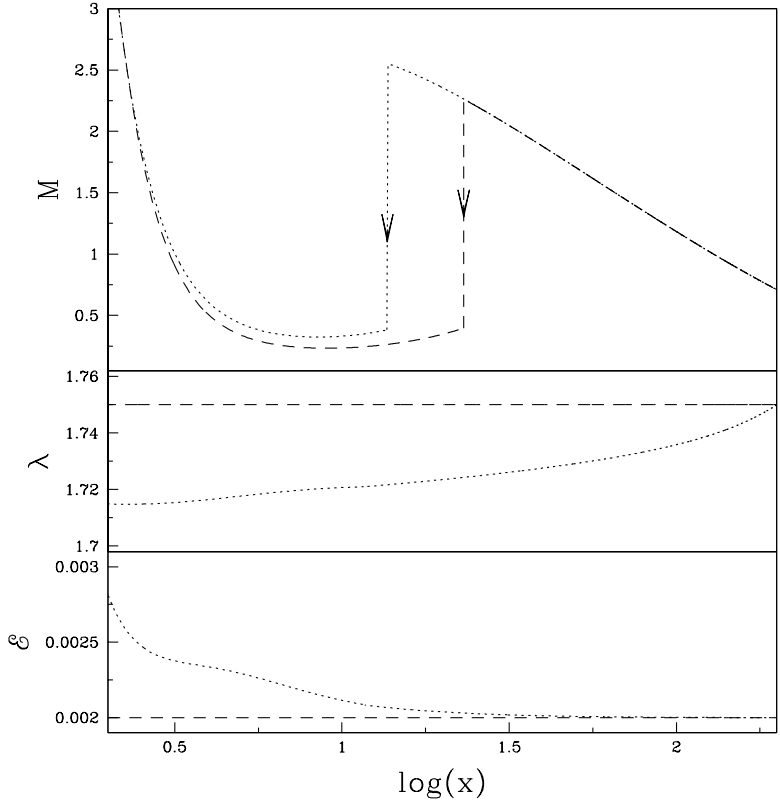}
\end{center}

\caption{Variation of $M$ (upper panel), $\lambda$ (middle panel),
${\mathcal E}$ (lower panel) with $log(x)$, for
${\alpha}_{\Pi}=0$ (dashed) and
${\alpha}_{\Pi}=0.003$ (dotted). The outer edge quantities
are $x_{\rm inj}=200$, $({\mathcal E}_{\rm inj},
\lambda_{\rm inj})=(0.002,1.75)$. The shocks are
$x_s=20.3$
(dashed) and $x_s=11.65$ (dotted).
}
\label{fig6}
\end{figure}

\begin{figure}

\begin{center}
\includegraphics[scale=0.6]{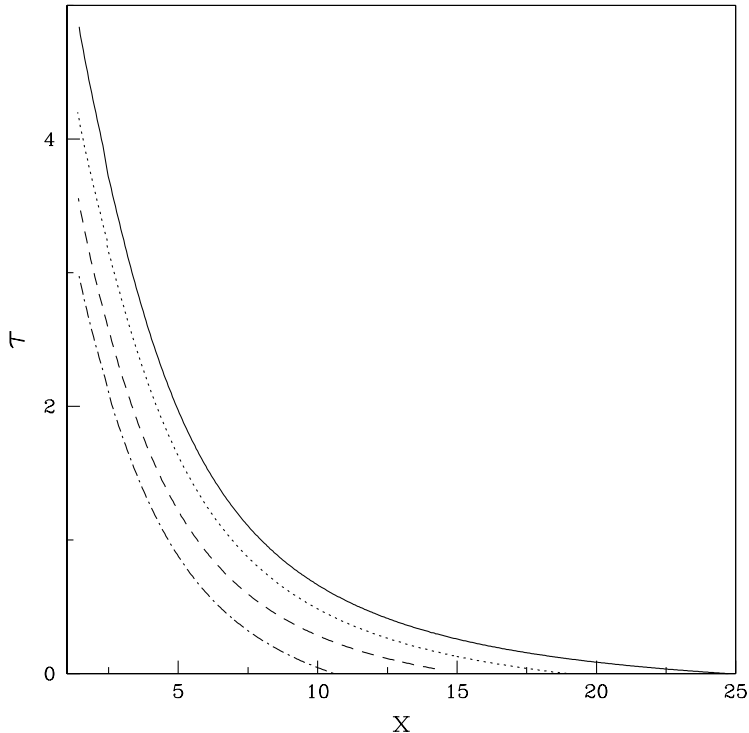}
\end{center}

\caption{Variation of $\tau$ with $(x)$, for
$({\alpha}_{\Pi}, R_{\dot m})=(0,0)$ (solid),
$({\alpha}_{\Pi}, R_{\dot m})=(0,0.093)$ (dotted),
$({\alpha}_{\Pi}, R_{\dot m})=(0.003,0)$ (dashed),
and $({\alpha}_{\Pi}, R_{\dot m})=(0.003,0.0.07)$
(dashed-dotted). The outer edge quantities
are $x_{\rm inj}=200$, $({\mathcal E}_{\rm inj},
\lambda_{\rm inj})=(0.002,1.75)$. Here $\tau$
is plotted for CENBOL.
}
\label{fig7}
\end{figure}

So far, we have shown that the mass outflow rates
strongly depends on shock location $x_s$, and formation of shock
itself depends crucially on the viscosity parameter.
We have also shown that shocks form closer to the black hole
for viscous flows than in inviscid accretion flow. However, we are yet
to comment on the observational consequence of such phenomena.
Admittedly, it is beyond the scope of the present paper to
compute the detailed spectrum of the flow, but we would like to make
some qualitative remarks on the spectral properties of such flows.
It is to be remembered, \citet{ct95} showed that hard power law photons were
generated at CENBOL by
inverse-Comptonization of soft photons from the pre-shock
region of the disc.
Thus we calculate the optical depth of the CENBOL, as this provides a
qualitative understanding of the nature of the spectrum.
The definition of the optical depth for photons entering
the CENBOL is given by,
$$
\tau (x)={\int}^x_{x_s}\kappa \rho dx^{\prime},
$$
where, $\kappa$ is the opacity assuming Thompson scattering cross-section,
and $\rho$ is the density of the flow.
We consider matter with identical outer boundary conditions
$(x_{\rm inj},{\mathcal E}_{\rm inj},\lambda_{\rm inj})=(200,0.002,1.75)$.
The mass accretion rate at $x_{\rm inj}$
is ${\dot M}_-=0.3{\dot M}_{\rm Edd}$, and the central mass is
$M_{BH}=10M_{\odot}$.
In Fig. 7, $\tau(x)$ is plotted
only for the post-shock disc with $({\alpha}_{\Pi}, R_{\dot m})=(0,0)$ (solid),
$(0,0.093)$ (dotted),
$(0.003,0)$ (dashed),
and $(0.003,0.07)$
(dashed-dotted), respectively.
For the most simplified case (solid), the shock is at $x_s=25$, and the optical
depth $\tau{\sim}1$ is at $8r_g$. Therefore most of
the soft photons entering the CENBOL fails to sample the inner part
of the disc.
In presence of mass loss (dotted), the density of the CENBOL is lowered
which reduces its optical depth.
For viscous flow (dashed), optical depth is much
lower compared to the inviscid flow (solid) as the shock forms ($x_s=14.2$)
closer to the black hole.
If we allow mass loss (dashed-dotted), the optical depth is reduced further.
Therefore, most of the
soft photons coming from the pre-shock region could interact with the
hot electrons
of the inner part of the disc ($\tau\sim 1$ at $x{\lsim}4r_g$)
and there by cool it.
A cooler CENBOL will evidently
produce softer spectrum compared to the inviscid case.
Chakrabarti (1998) calculated spectrum of a inviscid disc which
is loosing mass from the post-shock region, and indeed he found that the
spectrum softens, which correspond to solutions similar to the
dotted curve. From our self-consistent study, we predict that the twin effects
of viscosity and mass loss will soften the spectrum even further.

\section{Concluding Remarks}
Early studies
of transonic accretion flow
showed the existence of standing or oscillating shocks \citep{c89}.
The presence of shocks were
also thought to be the real cause behind hard state of the disc spectrum
\citep{ct95}, and also
generating outflows \citep{c99}. Studies of shocks in viscous accretion
disc
were also undertaken, but were confined
to obtain the global solution for accretion flows with or without
shocks and/or the detailed study of the parameter space
\citep{gut,cd04}. Important as they are, since a proper understanding
enhances the grasp on the physics of such flows, no attempts
have
been undertaken to link these viscous flows with observables
such as outflows and spectrum.
In this paper we have extended our earlier computation
of mass loss \citep{dcnc01}, by employing the knowledge of viscous
disc solutions
\citep{cd04}. In the earlier paper \citep{dcnc01}, the disc was considered
to be inviscid and the jet was non-rotating. Furthermore, the shock
condition was not modified for mass loss. 
Presently, we have included viscosity and considered 
rotating outflows, in particular we have
concentrated on studying how viscosity affects
shocks and in turn, how shocks affect the mass outflow rates.

We have shown that mass loss from the post-shock disc reduces post-shock
total pressure, and hence the shock front moves
towards the black hole. We have also shown that the mass outflow rate
depends on the energy, as well as the angular momentum
of the flow, and increasing both, in turn, increases the mass outflow rate.
However, less energetic accretion will produce lower mass outflow rates,
even though angular momentum is increased outward, by increasing the
viscosity parameter. 

More interestingly, we have shown that outflow rate decreases with the
increase of viscosity parameter for flows with same outer boundary conditions.
This is due to the weakening of centrifugal barrier with the
increase of viscosity parameter.
We have also shown that the post shock optical depth decreases in presence
of viscosity and mass loss, which may soften the emitted spectrum.
A detailed study of spectrum from such viscous discs,
by including cooling processes is under consideration
and will be reported elsewhere.

\ack

I. C. was supported by the KOSEF grant R01-2004-000-10005-0,
and S. D. was supported by KOSEF through Astrophysical Research Center
for the Structure and Evolution of the Cosmos (ARCSEC). The authors
also acknowledge the anonymous referee for his fruitful suggestions.

\end{document}